# Coherent multi-flavour spin dynamics in a fermionic quantum gas

Jasper Simon Krauser[1,2], Jannes Heinze[1,2], Nick Fläschner[1], Sören Götze[1], Christoph Becker[1] and Klaus Sengstock[1,*]

[1] Institut für Laserphysik, Universität Hamburg, 22761 Hamburg, Germany
[2] These authors contributed equally to this work.
* e-mail: klaus.sengstock@physnet.uni-hamburg.de

**Microscopic spin interaction processes are fundamental for global static and dynamical magnetic properties of many-body systems. Quantum gases as pure and well isolated systems offer intriguing possibilities to study basic magnetic processes including non-equilibrium dynamics. Here, we report on the realization of a well-controlled fermionic spinor gas in an optical lattice with tunable effective spin ranging from 1/2 to 9/2. We observe long-lived intrinsic spin oscillations and investigate the transition from two-body to many-body dynamics. The latter results in a spin-interaction driven melting of a band insulator. Via an external magnetic field we control the system's dimensionality and tune the spin oscillations in and out of resonance. Our results open new routes to study quantum magnetism of fermionic particles beyond conventional spin 1/2 systems.**

Magnetism plays a key role for the fundamental understanding of materials and in modern technologies. A major focus is the understanding of quantum properties of magnetism, which have their origin in the underlying microscopic processes between elementary spins. However, it remains challenging to derive macroscopic magnetic phenomena such as the fractional quantum Hall effect or the formation of spin liquids directly from the microscopic level. Here, scalable and controllable model systems allow to bridge this gap. Alongside a few tunable magnetic condensed matter systems[1,2], atomic physics experiments came into focus in the last years due to their unrivaled control over all experimental parameters and nearly perfect isolation from environmental influences. Ion chains[3,4] as well as bosonic quantum gases, either harmonically trapped or confined in optical lattices, have produced striking results towards the simulation of classical and quantum magnetism[5-14]. Lacking Pauli blocking, however, these experiments did not catch the fermionic character of electronic magnetism. In this direction, the possibility of itinerant ferromagnetism[15-18] and spin transport[19] in bulk fermionic quantum gases has been recently discussed. To resemble electron spins in real solids even better, it is desirable to have a fully controllable fermionic lattice quantum simulator. In addition, with fermionic quantum gases, completely new systems can be realized, e.g. high-spin systems[20-22] $(s > 1/2)$, for which complex quantum phases are theoretically predicted: This includes unconventional BCS superfluids[23,24], QCD-like color superfluidity[25], SU(N)-magnetism with alkaline earth atoms[26-29] and further multi-flavour systems[30-33].

In this article, we demonstrate the first experimental realization of a well-controlled fermionic spinor gas with interaction-driven spin oscillations. By properly choosing the initial spin states we can change the effective length of the atomic spin from 1/2 to 9/2. The control over the magnetic field allows to initialize and stop spin dynamics and to select the number of involved levels. We extract the microscopic interaction parameters and find excellent agreement with a two-particle model including all spin-dependent interactions. By tuning the depth of the optical lattice, we investigated the transition from on-site dominated to quantum many-body spin dynamics, where spins diffuse throughout the lattice producing highly entangled states. For this case, we observe a new form of melting of a band insulator.

## Principles of fermionic spin-changing collisions

We perform our experiments employing a fermionic spinor gas of $^{40}$K atoms in the $f = 9/2$ manifold. Initially, we prepare an equal mixture of two different spin states with typically $N = 4 \cdot 10^5$ atoms in an optical lattice, forming a large-scale band insulator[34-36] (diameter about 50 lattice sites) as depicted in Fig. 1a (for details see Methods).

In order to shed light on the microscopic collision processes, we first consider two fermionic atoms in different spin states $|m_1\rangle$ and $|m_2\rangle$ in the lowest spatial mode on an isolated lattice site, forming a two-particle state $|m_1, m_2\rangle = 1/\sqrt{2}(|m_1\rangle|m_2\rangle - |m_2\rangle|m_1\rangle)$. For spin 1/2 particles with solely two hyperfine states, analogous to

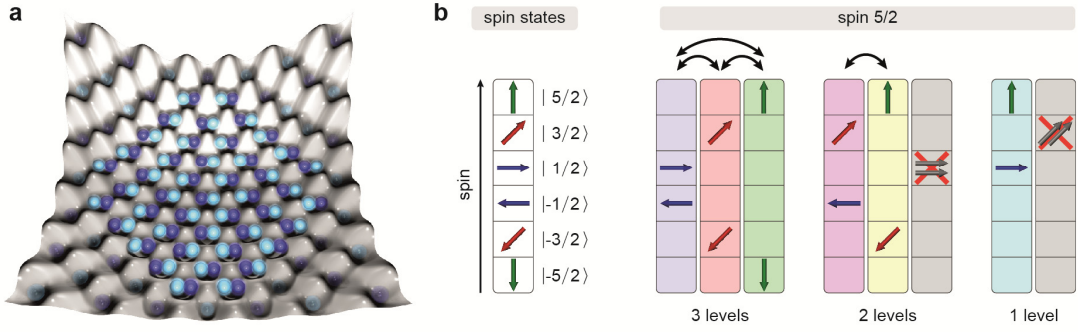

**Figure 1 | Principles of fermionic lattice spin dynamics. a**, Sketch of atoms in an optical lattice with harmonic confinement. A band insulator (with two particles per site) is formed in the center within the lowest spatial state. **b**, Microscopic collision processes on an isolated lattice site. Spin states for spin 5/2 particles are exemplarily shown (left). Colored boxes represent possible two-particle states (right). These states are coupled by spin-changing collisions, forming one-level, two-level and three-level systems. Magnetization conservation and Pauli blocking restrict the number of involved two-particle states. States represented by grey boxes are forbidden due to Pauli blocking.

electrons in solids, there is only one possible two-particle state. For high-spin particles $(s>1/2)$, spin-changing collisions[37-40] can transfer the atoms into new states $|m'_1, m'_2\rangle$ at low magnetic field. For increasing spin length, the number of involved two-particle states typically increases. Spin-changing collisions of fermionic atoms have to satisfy two physical restrictions: conservation of magnetization $M = m_1 + m_2 = m'_1 + m'_2$ and Pauli blocking. As an example, consider the case of spin 5/2 as shown in Fig. 1b. Depending on the initially chosen spin states, either a three-level, a two-level or a one-level system can be realized. This is a consequence of conservation of magnetization and Pauli blocking which restrict the number of allowed final two-particle states. In particular, all states with $m'_1 = m'_2$ are forbidden due to the Pauli exclusion principle. Employing $^{40}$K in $f = 9/2$ with its ten spin states allows to experimentally vary the number of involved two-particle states between one and five.

However, when finite tunneling couples neighboring lattice sites, these restrictions are significantly lowered. As depicted in Fig. 2, formerly forbidden final spin states can now be generated, leading to much more complex dynamics. For simplicity, we limit the explanation to three characteristic situations. First, consider two coupled lattice sites as shown in Fig. 2a and b, which are initially filled with the same two-particle state $|\nearrow, \rightarrow\rangle$, forming a band insulator (top). Due to Pauli blocking, tunneling is forbidden and thus the number of particles per site is fixed. Spin dynamics on each individual lattice site leads to a time-dependent occupation of the two-particle state $|\uparrow, \searrow\rangle$ and hence a finite probability that different two-particle states are realized on neighboring sites (upper middle). In this case, Pauli blocking no longer prevents tunneling. The dominating processes now depend on the lattice depth: For shallow lattices as shown in Fig. 2a, the system behaves like a metal. Single atoms can tunnel and induce local density fluctuations, e.g. one site contains one and the other site three particles (lower middle). This opens additional spin channels on the triply occupied site and leads to the formation of new spin states (bottom). For deep lattices, however, when the tunneling energy is small compared to the on-site energy, the system is in a Mott-insulating state with two particles per site and local density fluctuations are strongly suppressed. Further two-particle states can be realized only via super-exchange processes, as sketched in Fig. 2b (lower middle). Afterwards, the formation of new spin states is possible (bottom). Note, that in contrast to the situation of metallic tunneling, only the local magnetization fluctuates while the local density stays constant. In Fig. 2c, the scenario at edges or defects of the initial band insulator is shown, i.e. when a singly occupied site is adjacent to a doubly occupied site (top). After a spin-changing collision on the doubly occupied site (upper middle), a tunneling process can form another two-particle state on the initially singly occupied site (lower middle). Again, this leads to the realization of new spin states (bottom). In all of the above mentioned situations, the local magnetization is no longer conserved, which allows for the occupation of new spin states, which would be forbidden on perfectly isolated lattice sites. The combination of finite tunneling and spin-changing collisions thus interestingly leads to a continuous melting of the band insulator.

## Spin dynamics in deep optical lattices

In a first set of experiments, we studied the time evolution of exclusively doubly occupied sites (see Methods) in the initial two-particle state $|m_1, m_2\rangle = |1/2, 9/2\rangle$ in the low-tunneling regime. As one central result of this work, we observe long-lived coherent spin oscillations of fermionic atoms for the first time. As outlined above, the state $|1/2,9/2\rangle$ is only coupled to $|3/2,7/2\rangle$. Since the state $|5/2,5/2\rangle$ is forbidden due to Pauli blocking, this constitutes a two-level system. We observe oscillations over many periods for more than 250 ms with an amplitude in excess of 70 % as shown in Fig. 3a.

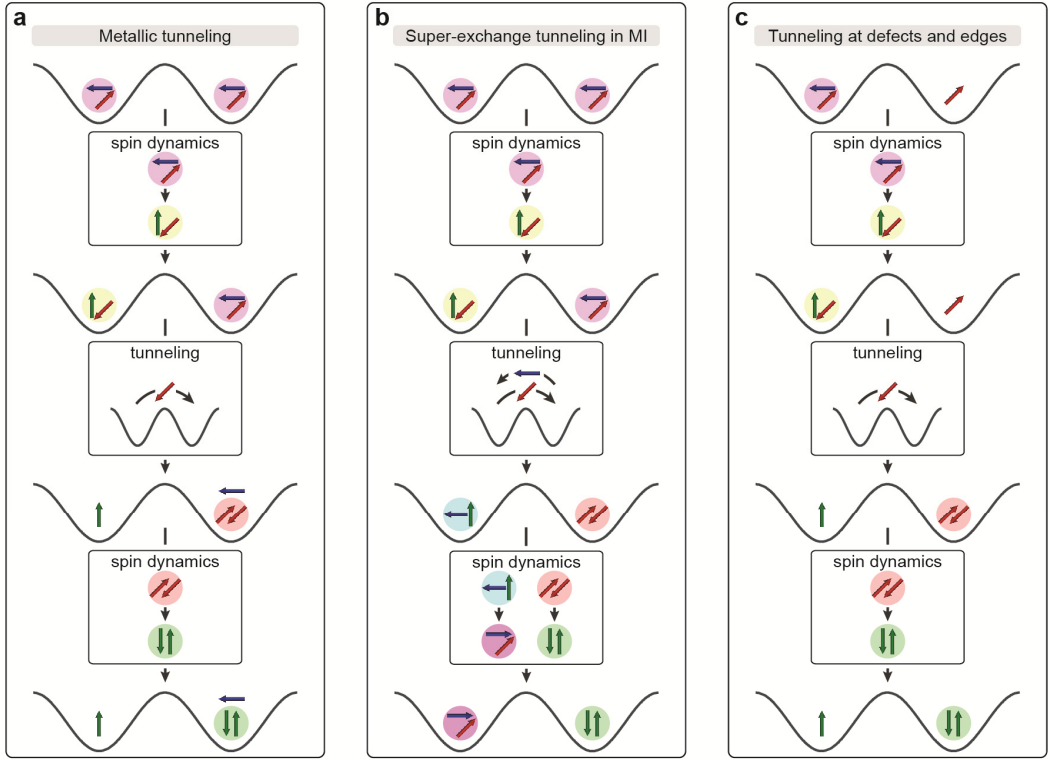

**Figure 2 | Simplified sketch of processes originating from the combination of spin-changing collisions and tunneling.** Shown are two lattice sites coupled by finite tunneling. Three characteristic processes are distinguished, each of them allowing for the generation of new spin states, which are forbidden on perfectly isolated lattice sites. **a**, Metallic tunneling occurs in the high-tunneling regime. Starting from uniform filling of each lattice site with the same two-particle state $|\nearrow,\rightarrow\rangle$, density fluctuations are induced. **b**, For large interaction strengths, the system is in a Mott insulating state with suppressed density fluctuations. Tunneling is only possible via super-exchange. **c**, Tunneling at edges and defects can always occur, independent of the ratio between tunneling and interaction energy.

Two energy scales are important for the spin-changing collisions as depicted in Fig. 3b: the difference in interaction energy and the difference in Zeeman energy between the individual two-particle states. At large magnetic fields, the atoms are pinned to a fixed spin by the Zeeman energy. For small magnetic fields, the spin interaction becomes relevant and spin dynamics occurs. The new eigenstates become quantum mechanical superpositions of the non-interacting two-particle states. A resonant feature appears in the spin oscillations when Zeeman energy and spin-dependent interaction energy are equal[41-43].

For the experiments, we first prepared the atoms in the Zeeman-dominated regime. Quenching the magnetic field to a lower value $B_{exp}$ initializes the observed spin oscillations. We investigated the crossover between the interaction-dominated and the polarized regime by studying spin dynamics at different $B_{exp}$. Extracting both the frequency and the amplitude of the oscillation, a clear Rabi-resonance feature as expected for a two-level system is observed and depicted in Fig. 3c. We identify the resonance position to be at $B_{exp} = 0.175 \pm 0.006\,\text{G}$, where the observed amplitude possesses a maximum while the frequency is minimal. For higher magnetic fields, the system approaches the polarized regime and the spin oscillations vanish. We compare our data to a two-particle model and find excellent agreement. Since the oscillation frequency at resonance is related to the difference of the scattering lengths $a_F$ in the respective scattering channels with total spin $F = 8$ and $F = 6$, our measurements provide a high precision test for molecular calculations of scattering lengths. We compare our measured value of $a_8 - a_6 = 2.26\,a_B$ to theoretical predictions using coupled-channel calculations and find very good agreement. Beyond these pure on-site effects, we observe a slow appearance of the spin states $|m\rangle=|5/2\rangle$ and $|-1/2\rangle$, which would be blocked at zero tunneling. In the regime of deep lattices, only super-exchange tunneling and tunneling at edges and defects can occur. We attribute the appearance of new spin states to the latter, as the timescale for super-exchange tunneling is much too long for our experimental parameters ($\Gamma^{-1}>10\,\text{s}$). We also observe a damping of the coherent spin oscillations at a time scale of $\Gamma^{-1}=96.4\pm4.6\,\text{ms}$, which we believe to be mostly due to edge and defect effects, but can also be influenced by magnetic field noise and gradients, which we have estimated, however, to be small.

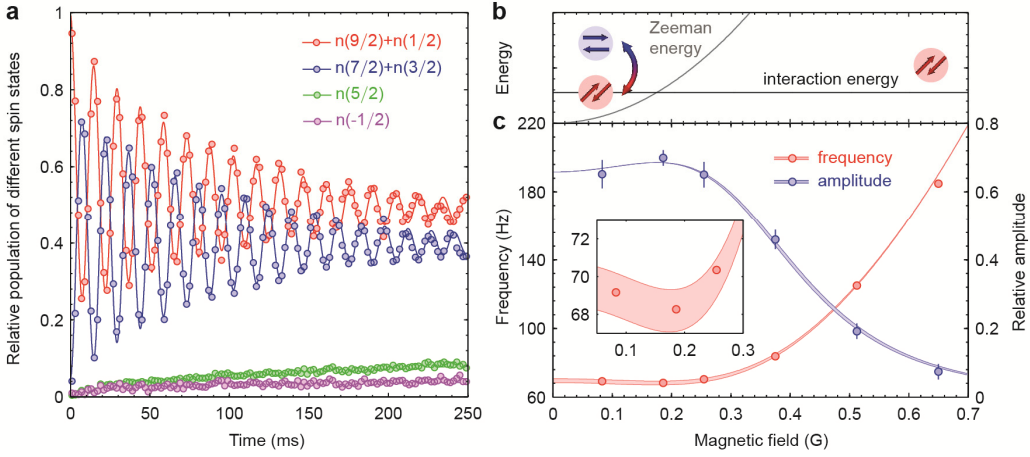

**Figure 3 | Coherent fermionic spin dynamics. a**, High contrast spin oscillations of a fermionic quantum gas. Plotted are the relative populations $n(m)$ of different spin states $|m\rangle$ as a function of time after initialization. Solid lines are fits to the data, from which we extract oscillation amplitude and frequency (for details see Methods). The lattice depth is $35\,E_r$ (with $E_r$ the lattice recoil energy) and the magnetic field is $B_{\text{exp}}=0.186\,\text{G}$. **b**, Sketch of the crossover from the interaction-dominated to the Zeeman regime. The spin resonance appears at the crossing of the two solid lines. **c**, Frequency and amplitude deduced from fits to spin oscillations as shown in Fig. 3a for different magnetic fields. Error bars correspond solely to fit errors, representing two standard deviations. The red and blue curves result from a two-body calculation. The widths of the curves are given by the lattice depth uncertainty. The inset shows a zoom into the spin resonance area.

In a second set of experiments, we realized a high-spin system by preparing the state $|1/2,-1/2\rangle$ (see Methods) which couples to the states $|3/2,-3/2\rangle$, $|5/2,-5/2\rangle$, $|7/2,-7/2\rangle$, and $|9/2,-9/2\rangle$. This constitutes an effective five-level system. In this situation, we observe complex multi-flavour quantum dynamics including all five two-particle states now governed by up to ten frequencies. This becomes apparent in Fig. 4a, where the corresponding beat-notes of the signal are clearly visible.

We have investigated the spin dynamics of this multi-flavour system at different magnetic field strengths and find that the amount of contributing oscillation frequencies increases with decreasing magnetic field (Fig. 4c). This is in good agreement with our numerical simulation of the complete multi-flavour dynamics, which reveals that the external magnetic field provides full control over the effective dimensionality of the system in spin space. The number of eigenstates significantly overlapping with the initial state depends crucially on the magnetic field as shown in Fig. 4b. Each observed frequency can be assigned to the superposition of two eigenstates. Typical Fourier spectra for the effective four-level system at a magnetic field of $B_{\text{exp}}=0.372\,\text{G}$ and for the effective two-level system at a magnetic field of $B_{\text{exp}}=1.014\,\text{G}$ are exemplarily shown in Fig. 4d. In this multi-flavour spin system with its rich dynamics, the observed damping is similar to the two-level case.

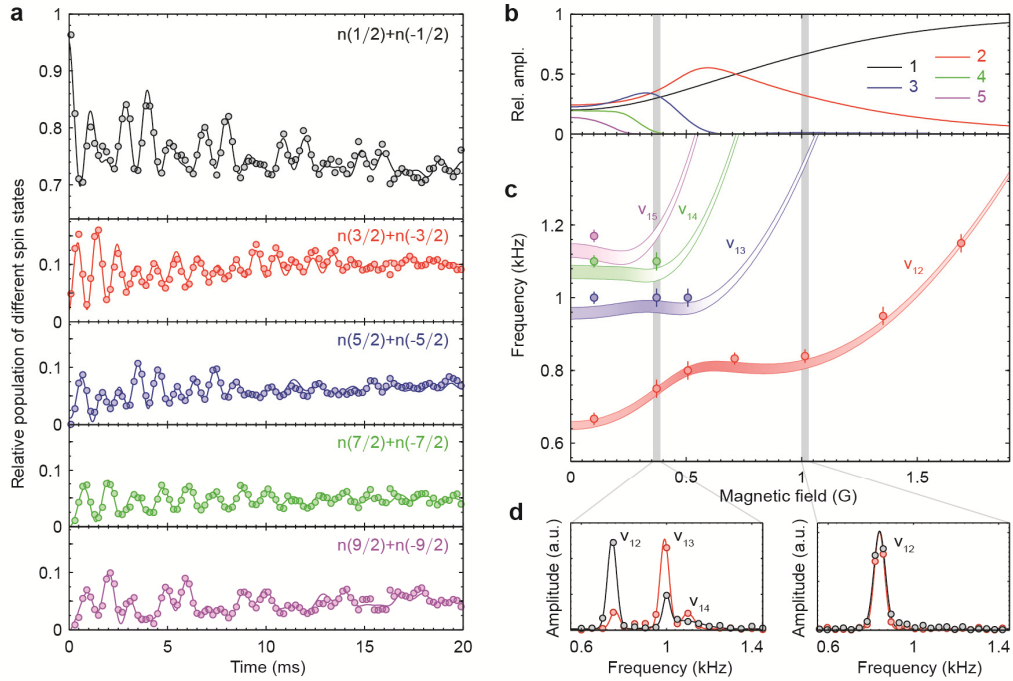

**Figure 4 | Coherent multi-flavour spin dynamics of fermionic atoms. a**, Coherent spin oscillations with five two-particle states involved. Plotted are the relative populations $n(m)$ of different spin states $|m\rangle$ versus time. The lattice depth is $25\,E_r$ and $B_{exp} = 0.372\,\text{G}$. At these parameters, four levels effectively participate in the spin evolution, leading to pronounced beat notes in the signal. Solid lines are fits to the data. **b**, Calculated amplitudes of the overlap integral $|\langle n|1/2,-1/2\rangle|^2$ between the initial state and all five two-particle eigenstates $n=1,\ldots,5$ versus the magnetic field. At high magnetic fields, the ground state coincides with the initial state. For lower magnetic fields, an increasing number of eigenstates overlaps and thus contributes to spin oscillations. **c**, Observed frequencies, obtained by discrete Fourier analysis of spin oscillation data like in Fig. 4a for different magnetic fields with corresponding errors. Curves show the numerically calculated frequencies. Their widths are given by uncertainties in lattice depth. Shading of the curves is proportional to the expected strength of the transition from Fig. 4b. For lower magnetic field, the number of observed frequencies increases, fully consistent with calculations. **d**, Typical discrete Fourier spectra for the populations $n(1/2)+n(-1/2)$ (black lines) and $n(3/2)+n(-3/2)$ (red lines) at magnetic fields $B_{exp}=0.372\,\text{G}$ (left) and $B_{exp}=1.014\,\text{G}$ (right).

## Crossover to the many-body regime

The experiments presented so far have been performed in the low-tunneling limit, in which spin dynamics is strongly dominated by local interactions on individual lattice sites. It is now particularly interesting to investigate the crossover to the tunneling-dominated regime, where the system approaches fermionic bulk spin dynamics. In order to investigate this crossover, we again start in the $|1/2,9/2\rangle$ effective two-level band insulating system and then decrease the lattice depth along one spatial dimension to a final value $V_L$. As a striking result, high contrast spin oscillations with a frequency consistent with our pure two-particle model are also observed in this regime (see Fig. 5a). Note, that the on-site density and thus the frequency changes with lattice depth as expected (see Fig. 5b). Besides the change in frequency, we observe an enhanced damping with decreasing lattice depth up to a complete disappearance of the spin oscillations for very shallow lattices. At the same time, the population of the spin states $|5/2\rangle$ and $|-1/2\rangle$ increases. Considering the three tunneling scenarios presented in Fig. 2, we compare the expected timescales of these processes to the measured damping constants. For super-exchange tunneling the expected timescale is proportional to $h^2\nu/J^2$, with $J$ the single-particle tunneling energy in the weak lattice direction and $\nu$ the spin oscillation frequency. For tunneling at edges, the expected damping time depends on the propagation time per site which is on the order of $h/J$ [44]. A decrease of the global spin oscillation amplitude to $1/e$ would require a time of $hd(e-1)/(4Je)$ with $d$ the length of the quasi 1d tube along the weak lattice direction. Especially at low lattice depths, the damping cannot be explained solely by these processes. Since the observed damping rate is on the order of the tunneling rate $J$, we attribute the enhanced damping as well as the increased population of initially forbidden spin states to metallic tunneling in the core of the system as described in Fig. 2a. The combination of tunneling and spin-changing collisions dynamically opens up new spin channels and provides a pathway towards a multi-flavour Fermi system, which is no longer limited to two spins per lattice site. In this situation, we expect increasing local fluctuations of the magnetization in contrast to the fixed local magnetization in the initial band insulator. Regarding the many-body character of the system, we interpret this process as an intrinsic spin-collisionally induced melting of the band insulator.

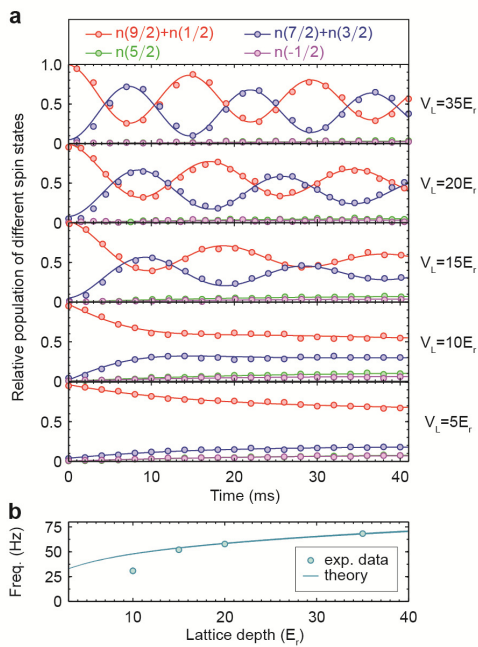

**Figure 5 | Transition from on-site to many-body spin dynamics. a**, Relative populations $n(m)$ of different spin states $|m\rangle$ as a function of time. Solid lines are fits to the data, from which we extract oscillation amplitude and frequency. The magnetic field is $B_{\text{exp}} = 0.186\,\text{G}$. The lattice depth $V_L$ along one dimension is indicated on the right hand side, for the other two dimensions the lattice depth is $35\,E_r$. The transition from single-site dominated high contrast oscillations to diffusing spins in the high-tunneling regime is clearly visible with decreasing lattice depth. In the latter case, spin oscillations in the initial channels are strongly damped, while at the same time tunneling leads to the occupation of new spin states. **b**, Extracted oscillation frequencies from Fig. 5a compared to a two-particle calculation (solid line). Errorbars representing two standard deviations lie within the datapoints. The width of the curve is given by the lattice depth uncertainty.

In conclusion, we have thoroughly studied a high-spin fermionic spinor gas in an optical lattice. Our results reveal the complex dynamics of these systems governed by the interplay between spin-dependent interactions, Zeeman energy, Pauli blocking and tunneling. For the first time, we have observed long-lived coherent spin oscillations of fermionic atoms. We have demonstrated the ability to fully control these spin oscillations, which allowed us to identify a clear spin resonance feature and to extract microscopic interaction parameters with high accuracy. Our results show excellent agreement with a two-body calculation in deep optical lattices. Hence, they provide a promising starting point towards several new scientific directions: Fermionic spinor gases in optical lattices are ideally suited to study magnetic many-body properties on a microscopic level, e.g. spin transport, spin-charge separation, spin waves or spin-orbit coupling. In the strongly-correlated regime, spin dynamics could serve as an intriguing tool to directly probe super-exchange processes. The tunability of the spin length opens up the route to further investigate high-spin magnetism in comparison to conventional spin 1/2 electrons and shows the ability to access fundamentally new multi-flavor spin systems. Especially for high-spin systems, our experiments also pave the way for coherent control of entanglement allowing studying quantum coherence properties.

## Acknowledgments

We acknowledge stimulating discussions with E. Demler, A. Eckardt, T.-L. Ho, M. Lewenstein and M. W. Zwierlein and thank T. Hanna and L. Cook for providing us calculated values of the scattering lengths. This work has been funded by DFG grant FOR 801.

## Methods

**Preparation of a fermionic quantum gas in an optical lattice.** By sympathetic cooling with $^{87}$Rb, about $N = 2 \cdot 10^6$ spin-polarized $^{40}$K atoms are cooled to quantum degeneracy in a magnetic trap. The potassium atoms are transferred adiabatically to a crossed optical dipole trap operated at $812\,\text{nm}$. After switching off the magnetic trap, a series of rf-pulses and sweeps is applied to prepare an equal mixture of two hyperfine states. The atoms are evaporatively cooled by exponentially ramping down the trap laser intensity within $2\,\text{s}$. The resulting spin mixture now consists of about $N = 4 \cdot 10^5$ atoms at typical temperatures between $0.15\,T_F$ and $0.25\,T_F$. For the presented experiments, the magnetic field after the evaporation is $7\,\text{G}$ for the mixture $|1/2\rangle + |-1/2\rangle$ and $3\,\text{G}$ for $|3/2\rangle + |7/2\rangle$. To increase the chemical potential, we ramp up the depth of the optical dipole trap in 50 ms and obtain final trapping frequencies of $\omega_x = 2\pi \cdot 125\,\text{Hz}$ in the vertical direction and $\omega_y = 2\pi \cdot 41\,\text{Hz}$ and $\omega_z = 2\pi \cdot 32\,\text{Hz}$ in the horizontal plane. Afterwards, we adiabatically ramp up a 3d cubic optical lattice in 150 ms. It consists of three orthogonal retro-reflected laser beams at 1030 nm with a $1/e^2$ radius of 200 μm, detuned with respect to each other by several 10 MHz. The uncertainty of our lattice depth calibration is $\pm 2\,\%$. The lattice depth and chemical potential are chosen such that a band insulator is formed in the center of the system which contains about 40 % of the atoms which we infer from the measured double occupancy. Note that only at low enough temperatures a sufficiently pure two-component band insulator is formed in the core of the lattice, which is crucial to observe high contrast spin oscillations. After the preparation procedure, we initialize spin dynamics by rapidly decreasing the magnetic field to a value $B_{\text{exp}}$ with an accuracy of about $\pm 0.003\,\text{G}$.

**Microwave transfer of singly occupied sites.** Atoms on singly occupied sites do not participate in the spin-changing dynamics and only contribute as a constant offset to the signal. We circumvent this by applying microwave pulses to transfer all singly occupied sites to the $f = 7/2$ hyperfine manifold, where the atoms are not resonant with our detection light (the $f = 7/2$ states are separated by a frequency of 1.3 GHz from the $f = 9/2$ states). For this, we prepare the atoms in the state $|3/2,7/2\rangle$ and let the system evolve on resonance for one quarter of a spin-oscillation period. Now, the doubly occupied sites are in the state $|1/2,9/2\rangle$. At this moment we stop the dynamics, by ramping up the magnetic field to $B = 1.69\,\text{G}$. Then, we apply four linearly polarized microwave pulses with a length between 35 μs and 50 μs and transfer all atoms on singly occupied sites in the remaining spin states ($|-1/2\rangle$, $|3/2\rangle$, $|5/2\rangle$ and $|7/2\rangle$) into the $f = 7/2$ manifold. Subsequently, we restart the spin dynamics, now from a very pure state, where all atoms contributing to the signal are in the two-particle state $|1/2,9/2\rangle$. Thus, we observe high contrast spin oscillations. In addition, the admixture of new spin components due to defect tunneling is strongly suppressed, since atoms in $f = 7/2$ do not exchange their spin with atoms in $f = 9/2$. Note, that this procedure was applied for the measurements starting from the initial state $|1/2,9/2\rangle$ in Fig. 3 and Fig. 5.

**Counting of spin components.** After a variable evolution time, we stop the spin oscillations and detect the relative occupations in the different states $|m\rangle$. To stop the oscillations we increase the magnetic field rapidly, which pins the atoms to their momentary state. Afterwards, we ramp down the lattice potential in 500 μs and hence map the quasi momenta of the atoms onto real momenta. After all optical potentials are switched off, we perform a Stern-Gerlach separation of the different spin states within a time-of-flight of 18.5 ms. The atoms are finally detected via absorption imaging with a short pulse of resonant laser light. We count the number of atoms in all spin states, accounting for saturation effects due to the spatial variation of the pulse intensity.

**Fitting procedure of the spin oscillations.** For each experiment, we measure the time-dependent occupation $n^{(m)}(t)$ of ten different single-particle spin states $|m\rangle$. To each of the spin states showing an oscillatory behavior we fit:

$$n^{(m)}(t) = A^{(m)} + B^{(m)} e^{-\Gamma_t^{(m)} t} + \sum_j C_j^{(m)} e^{-\Gamma_j t} \cos(\omega_j t + \varphi_j).$$

Here, $A^{(m)}$ is a general offset. The slow overall increase or decrease of the population of the individual spin states is expressed in the second term, employing an exponential function with amplitude $B^{(m)}$ and time constant $\Gamma_t^{(m)}$. The spin oscillations are described by the sum in the third term: $C_j^{(m)}$ is the amplitude, $\omega_j$ the frequency, and $\varphi_j$ the initial phase of the oscillation. We assume a damping of the oscillations described by an exponential function with the time constant $\Gamma_j$. The sum is taken over the number of all contributing frequencies $j$ as expected from theory. Note, that the values $\omega_j$, $\varphi_j$ and $\Gamma_j$ for each contributing frequency $j$ are extracted in conformity with all participating single-particle spin states.

For the two-level system (see Fig. 3), four single-particle spin states are involved for the on-site dynamics ($|1/2\rangle$, $|3/2\rangle$, $|7/2\rangle$, $|9/2\rangle$). Here, the switching time of the magnetic field is small compared to the observed oscillation period, which allows us to set the initial phases $\varphi_j = 0$. Further spin states, which are slowly populated due to tunneling processes, do not show coherent oscillations, but a slow overall increase of their population. We fit the occupation of these spin states with an exponential function $n^{(m)}(t) = A^{(m)}(1 - e^{-\Gamma_t^{(m)} t})$. Again, $A^{(m)}$ is the amplitude and $\Gamma_t^{(m)}$ the time constant.

For the investigated five-level system (see Fig. 4), five two-particle states ($|1/2,-1/2\rangle$, $|3/2,-3/2\rangle$, $|5/2,-5/2\rangle$, $|7/2,-7/2\rangle$, $|9/2,-9/2\rangle$) participate in the spin evolution involving all ten spin states ($|-9/2\rangle$, ..., $|9/2\rangle$). In this case, the oscillations are slightly influenced by the finite switching time of the magnetic field (500 μs), which we account for by including individual phases $\varphi_j = 0$ for each oscillation. The fit results coincide with the values of the discrete Fourier transform given in the paper. Note, that all fitting errors are given by the 95 % confidence-interval.

**Fitting procedure to extract the scattering length difference.** In Fig. 3c, the observed frequencies and amplitudes of the two-level system are depicted. The frequency depends on the magnetic field as well as the scattering energy difference of the participating channels. Since the lattice depth and the magnetic field are known precisely, the largest uncertainty originates from the scattering length difference $a_8 - a_6$. We fit the dependence of the frequency on the magnetic field $\nu(B)$ to the data for $B_{\text{exp}}$ up to 0.6 G, where the spin-dependent interaction plays a crucial role. We obtain a scattering length $a_8 - a_6 = 2.26\,a_B$ with a statistical error of $\pm 0.02\,a_B$ and systematic errors due to lattice depth ($\pm 0.04\,a_B$) and magnetic field uncertainties ($\pm 0.01\,a_B$).

This is in good agreement with the theoretically predicted value $a_8 - a_6 = 168.53\, a_B - 166.00\, a_B = 2.53\, a_B$. To adapt for a total spin oscillation amplitude smaller than unity, we rescale $A(B)$ by a fitted factor $\alpha = 0.69 \pm 0.02$ in Fig. 3c.